\documentclass{article}
\usepackage{sao2}
\usepackage{psfig}
\def\secdot{\hbox{$.\!\!^{\prime\prime}$}}  % Fractions of arcseconds
\def\magdot{\hbox{$.\!\!^{m}$}}

\begin{document}
\markboth{Barsukova, Fabrika, Pustilnik, Ugryumov}
{Optical monitoring of CI\,Cam}
\title{{\normalsize \it Bull.\,Spec.\,Astrophys.\,Obs.,}{\normalsize
\rm 1998, 45, 147-153}\vspace{0.4cm} \newline
{\LARGE \bf NEWS}\vspace{0.4cm}\newline
Optical monitoring of CI\,Cam after
the X--ray burst\\
on April 1, 1998}
\author{
E.A.\,Barsukova, S.N.\,Fabrika, S.A.\,Pustilnik, A.V.\,Ugryumov}
\institute{\saoname}
\date{December 14, 1998}{December 16, 1998}
\maketitle

\begin{abstract}
Results of optical spectral monitoring of CI\, Cam just after its a unusual bright and
rapidly evolving X-ray burst  are reported. This star is
a symbiotic-type X-ray binary.
The X-ray--radio--optical burst and the appearance of relativistic
S-shaped SS433-like jets make CI\, Cam to be a very interesting star.
We  observed emergence of a high excitation and variable
 emission line spectrum,  with broad components of hydrogen and
helium lines and the nebular forbidden line of NII.
All  this agrees well with  the formation of
relativistic jets in the burst.
It was concluded that the burst  resulted in destruction of external
stationary accretion disk regions.
The peak X-ray luminosity
is $\sim 10^{37}$~erg/s at  the adopted distance of 1~kpc. It is
well below  the Eddington luminosity, which is necessary for
 jet production in
a supercritical accretion disk. We propose  that jets were formed in
a collimated
accretion--ejection process  in a strong magnetic field of a neutron star.
\end{abstract}
\section*{X-ray burst on April 1}

 The bright and rapidly rising transient X-ray source XTE\,J0421+560
was detected
by the RXTE observatory in All Sky Monitor (ASM) observations on March 31, 1998
(Smith et al., 1998). Having a flux less than 40~mCr in  2--12~keV
band  on
March 31.36 UT, this source has brightened up to 2~Crabs on April 1.04
 according to  the RXTE data. The CGRO/BATSE ASM
 observed  even a more rapid
brightening of this source in  20--100~keV  band~--- from less than
55~mCr on March 31.03~-- 31.84 to 470~mCr on March 31.91
(Harmon et al., 1998; Paciesas and Fishman, 1998).
Such an extremely fast rise for a few hours is very unusual. Moreover
it was followed by a very fast fading (Fig.~1).
Both the BeppoSAX (Orlandini et al., 1998) and ASCA (Ueda et al., 1998)
monitoring observations on April 3--4
 showed the X--ray flux to decrease  40--50 times, and a gradual flux
fading was detected with an e--folding decay time of about 1 day.

\begin{figure}
\centerline{\psfig{figure=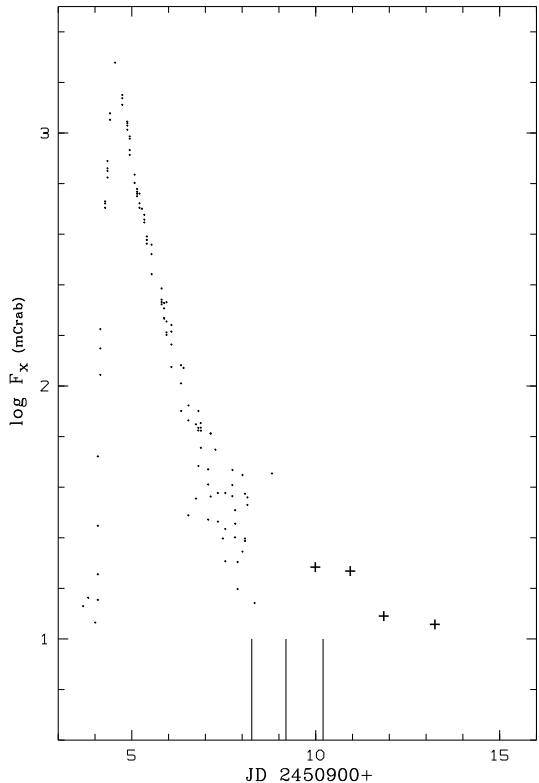,width=8cm}}
\caption{
The X-ray light curve of the April 1, 1998 burst of CI\,Cam from the
RXTE ASM data obtained in 2--10~keV  band.
(htpp://space.mit.edu/XTE/asmlc/srcs/data). Points mark the time-dwell
fluxes in mCrabs obtained with exposures of about 90 s. Crosses mark
daily-averaged data. Vertical bars mark the first three dates of the
spectral optical observations  described in this paper.}
\end{figure}

Beginning  from April 1.9 a radio  counterpart of XTE\,J0421+560 was discovered
in VLA observations (Hjellming and Mioduszewski, 1998a) at a level of about 12~mJy at 1.4~GHz.
The source  brightened rapidly and reached a level of 120~mJy
at 1.4~GHz on April 3.83 (Hjellming and Mioduszewski, 1998b).  This date the source size
was less than 0\secdot1 at 22.5~GHz. The spectrum slope and the
rapid variability  by factors of about 6, 3 and 2.3 at
1.4, 4.9 and 8.4~GHz, respectively (Hjellming and Mioduszewski, 1998b), indicated that this is
 synchrotron radio emission.
A ``core plus jet'' radio source was detected on the VLA  image  obtained
on April 5.08  (Hjellming and Mioduszewski, 1998c).
 Basing on VLA images  of this object, Hjellming and Mioduszewski
(1998c)  have reported  that beginning from April 5.08 the
source was resolved, and the extended emission  became apparent.
It had the appearance of a symmetrical S-shaped twin-jet,
strikingly similar to the radio jets of SS\,433.
The  apparent velocity of the outermost pair of emission components
 was found to be very high,
on average of 54~mas/day. The SS\,433-like rotating
corkscrew emission  pattern was fitted with a velocity in the range 0.3\,c to
0.4\,c at the  adopted distance of 1~kpc.  This SS\,433 jet-like
 mode of behaviour  distinguishes this transient X-ray star as a
very important target for relativistic stellar astrophysics.
The  derived jet velocity  evidences that the source of jets is
a relativistic  object --- a neutron star or a black hole.

This X-ray  radio source has been identified by Wagner and Starrfield
(1998) with the well-known symbiotic star CI\,Cam ($=$~MWC\,84)
(Merrill, 1933; Chkhikvadze, 1970; Downes, 1984; Bergner et al., 1995;
Miroshnichenko, 1995).
It was also considered to be a B[e] star. The identification
was  proved by
Wagner and Starrfield (1998) spectrum on April 3.145, where the strong HeII\,$\lambda$\,4686
 line was detected. Otherwise the spectrum was similar to that obtained
in January 1984 by Downes (1984) and described previously by
Merrill (1933). The HeII line had  not been detected at all in the previous studies.
Another photometric and spectroscopic study (Garcia et al., 1998)
was carried out on April 3.08--3.17. Garcia et al. have confirmed
the spectrum similarity to that of Downes (1984), but the  intensity of
FeII and HeI emisson lines has increased in comparison  with
hydrogen Balmer lines. The forest of strong H, HeI, FeII emission
lines obscured the continuum to the extent  that it is unclear if any
photospheric absorption lines were present. Garcia et al. (1998) found
${\rm V \approx 9\fm 2, B \approx 10\fm 2}$, which implies a brightening
of about 2\magdot4 and 2\magdot3,  respectively in comparison with the
pre-burst level (Bergner et al., 1995; Miroshnichenko, 1995). The usual star
brightness in V was $\approx 11\magdot6$ and in B $\approx 12\magdot5$.
It was quite
surprising that during the object's burst and brightening the colour
index did not change. UBVRI photometry obtained by Hynes et al. (1998)
on April 3.87  showed that the star faded by 0\magdot5 in V and B bands
 during 18 hours. The index B--V  was again the same as
 that in a quiescent state of the object (Bergner et al., 1995).

\begin{table}
\small
\caption{The journal of observations}
%\vspace{0.1cm}
\begin{tabular}{lrcrcr}
\rule{0pt}{0.02cm}\\
\hline
\rule{0pt}{0.02cm}\\
Date&{JD$^a$}&$\lambda\lambda$&$\Delta{\lambda}$&S/N&N\\
     & & \AA&\AA& & \\
\hline
\rule{0pt}{0.1cm}\\
1998 Apr 4&  08.26&   3800--6100&   4 &       40&    2\\
	  &&  5000--7400&    &         &    1\\
	 &&&&&      \\
1998 Apr 5&   09.19&  3800-6100&   4 &       20&    1\\
	  &&   5000-7400 &   &         &   1\\
	 &&&&&      \\
1998 Apr 6&  10.20&   3800-6100&   4 &       50&    2 \\
&   &   5000-7400 &   &  &           3   \\
	 &&&&&      \\
1998 Apr 19&   23.25  & 3800-6100&   7 &       30&   2     \\
&   &  5000-7400&    &  &           2       \\
	 &&&&&      \\
1998 May 16&   50.30&   3800-6100 &  4 &       50 &   2         \\
&  &   5000-7400&   & &           2           \\
\rule{0pt}{0cm}\\
\hline
\rule{0pt}{0.03cm}\\
\multicolumn{6}{l}{
[$^{a}$]\,\, JD (2,450,900.00 +)}
\end{tabular}
\end{table}

\section*{Observations}

The observations were carried out with the 6\,m telescope of the Special
Astrophysical Observatory (SAO RAS)
on April 4, 5, 6, 19 and May 16, 1998. The medium resolution
spectrograph SP--124  was used at the Nasmyth-1 focus
 with the UAGS camera and Photometrix CCD PM1024 ($24\times24\,\mu$m pixel size)
in all observations.
The spatial scale along the slit was equal to $0\farcs5$/pixel, the length of the slit
was about $40\arcsec$, the slit width was set to $1\arcsec$.
The grating B1 with 600 grooves/mm  produced the scale along the
dispersion 2.4 \AA/pixel with a spectral resolution of approximately
4 \AA\, (FWHM) in the spectral region from 3700 to 7500 \AA.
For data acquisition the system NICE under MIDAS (Kniazev and Shergin, 1995)
 was used.

\begin{figure*}
\centerline{\psfig{figure=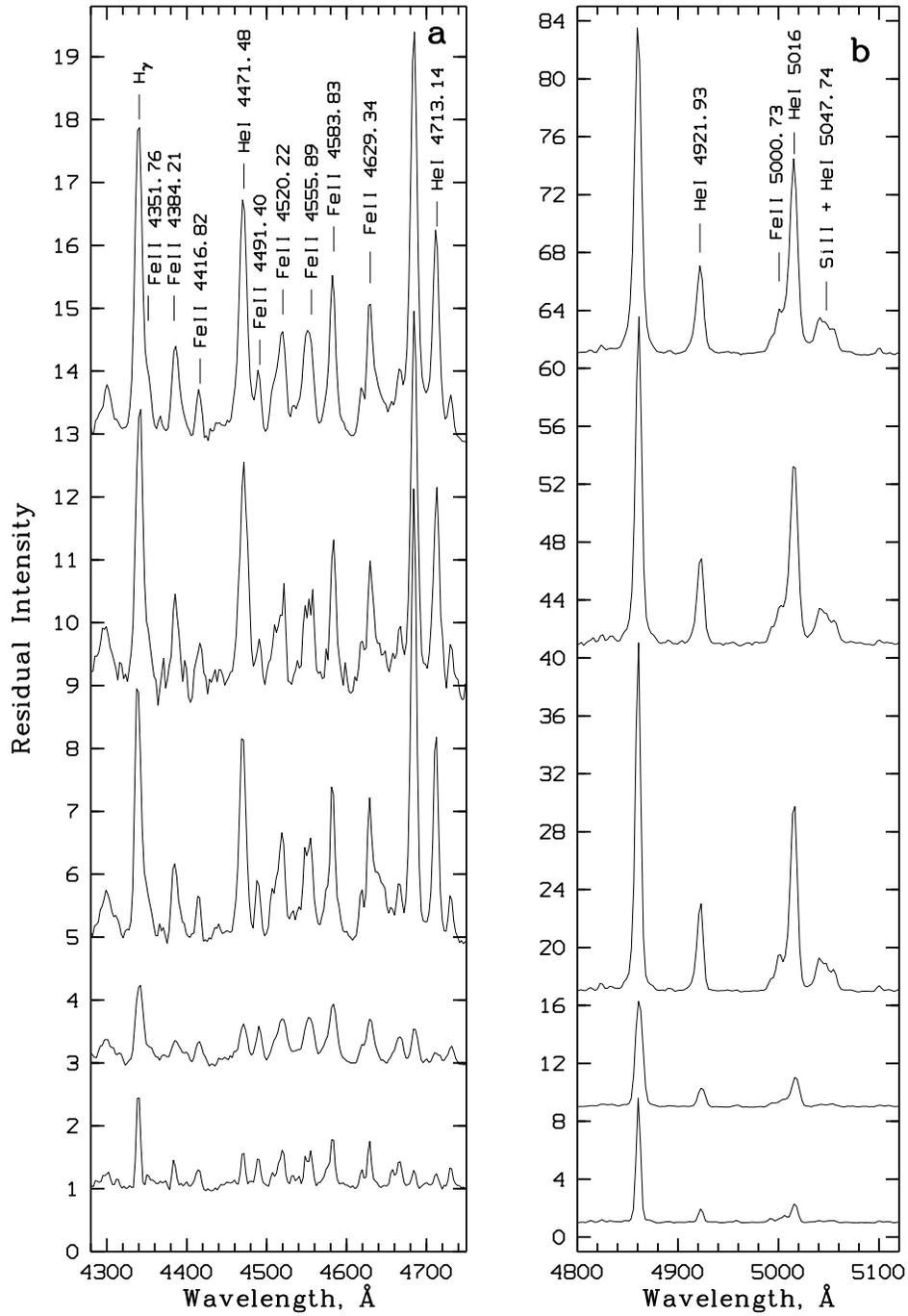,width=14cm}}
\caption{{\bf a--d.}
Fragments of normalized spectra obtained on April 4, 5, 6, 19 and
on May 16 (from top to bottom). The strongest lines in the
fragments~---   HeII~$\lambda\,4686$, H$\beta$,
FeII~$\lambda\,5169$, FeII~$\lambda\,5317$ and HeI~$\lambda\,5876$~---
are not marked.
 The spectra were accordingly shifted by:
a,c) 12, 8, 4, 2 and 0  units; b,d) 60, 40, 16, 8 and 0.}
\end{figure*}

\begin{figure*}
\setcounter{figure}{1}
\centerline{\psfig{figure=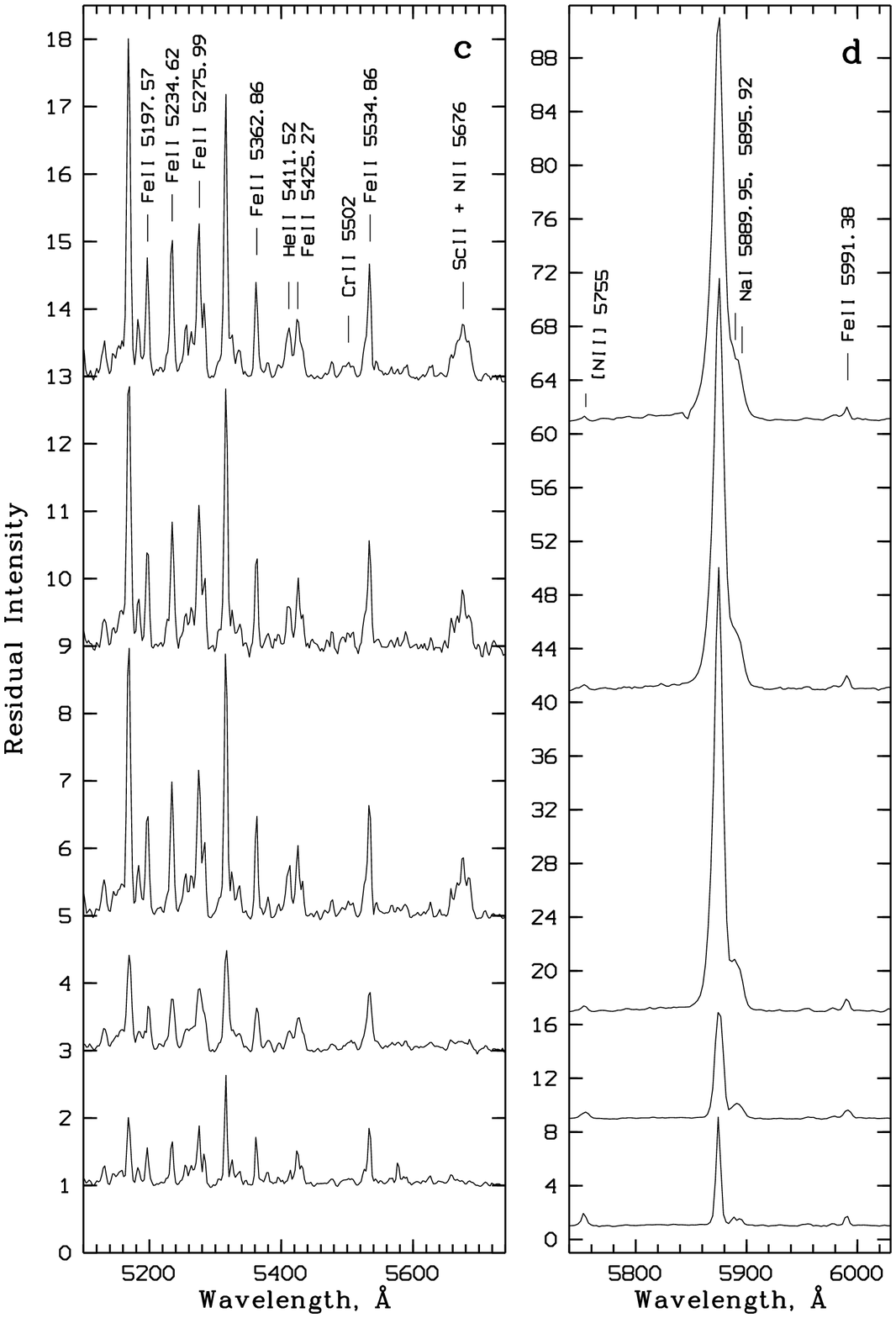,width=14cm}}
\caption{\bf c, d.}
\end{figure*}

All spectra  were reduced using the MIDAS context LONG (95NOV version)
adopted to perform automatic mode  reduction.
Primary 2--D CCD images were converted to 1--D spectra after  correction for the dark noise, debiasing and
sky subtraction.
 He--Ne--Ar source was used  to transform
2-D spectra to a linear wavelength scale.
For the dates April 4, 5, 6 spectrophotometric standards were observed, the data were
flux-calibrated and corrected for atmospheric extinction.

The  summary of observations ---  wavelength range,  spectral resolution,
 S/N ratio of individual spectra and  number of spectra are shown in Table 1.

\section*{Results}

The first   observations were carried out on April 4.75,1998, $\approx$1.6 days
 after the
observations by Wagner and Starrfield (1998) and Garcia et al. (1998),
 and accordingly 3.7 days after the X-ray burst maximum.
The moments of  our observations are marked in Fig.~1 by vertical bars.
We  show only our first three dates as after  the date
JD\,2450913 the X-ray flux was at  the level of measurement
 errors.

 Our spectral observations were made when the  object
X-ray flux faded at least by two orders of magnitude. The X-ray light curve
was obtained in the RXTE ASM observations (see
 htpp://space.mit.edu/XTE/asmlc/srcs/data).

 For each of the dates the observations consist of one or two
 spectra in red and blue wavelength range with an exposure time from 5 to 10 minutes.
The data on April 19, 1998 consist, in addition to these ordinary
mode spectra,  also
of 30 ``time-resolved'' 10-second  exposure spectra.  An analysis
of the rapid spectral variability of CI\,Cam will be presented later
elsewhere.
 Here we present the overall description of the spectrum and its
variability from date to date.

 In Fig.\,2  fragments of normalized spectra obtained on April
4, 5, 6, 19 and May 16 (from top to bottom) are presented. All the spectra in each
separate wavelength range are shown on the same scale for convenience.
It is obvious that the
spectrum excitation decreases with time. The HeII\,$\lambda 4686$
emission line (Fig.\,2a), being strong  in the beginning of April,
practically  has disappeared  in the May 16 spectrum.  The
same is correct for HeII~$\lambda 5411$. Note that the spectrum
on April 19  is of about twice worse resolution than  the spectra
on other dates.
HeI line intensities decrease faster than those of hydrogen lines.

The strong variability of SiII\,$\lambda 5041, 5056$ + HeI\,
$\lambda 5048$ + CII\,$\lambda 5045$ blend (Fig.\,2b)
is clearly seen. This blend  seems to disappear in the April 19 and
May 16 spectra. We estimate  that the contribution of the
SiII\,$\lambda 5041,\,5056$ lines is about 80\,\% of the flux in this blend,
while the HeI\,$\lambda 5048$ line contributes no more than 20\,\%.
 So the fast fading in intensity of this blend is caused by the
behaviour of the SiII lines.
The similar variability is seen in Fig.\,2c for the blend
ScII\,$\lambda 5658,\,5667,\,5669,\,5684$ + NII\,$\lambda 5676$ +
FeII\,$\lambda 5658$. For this blend we estimate  that about
50\,\% of the flux is from the NII line and about the same fraction is
in the ScII lines.

\begin{figure}
\centerline{\psfig{figure=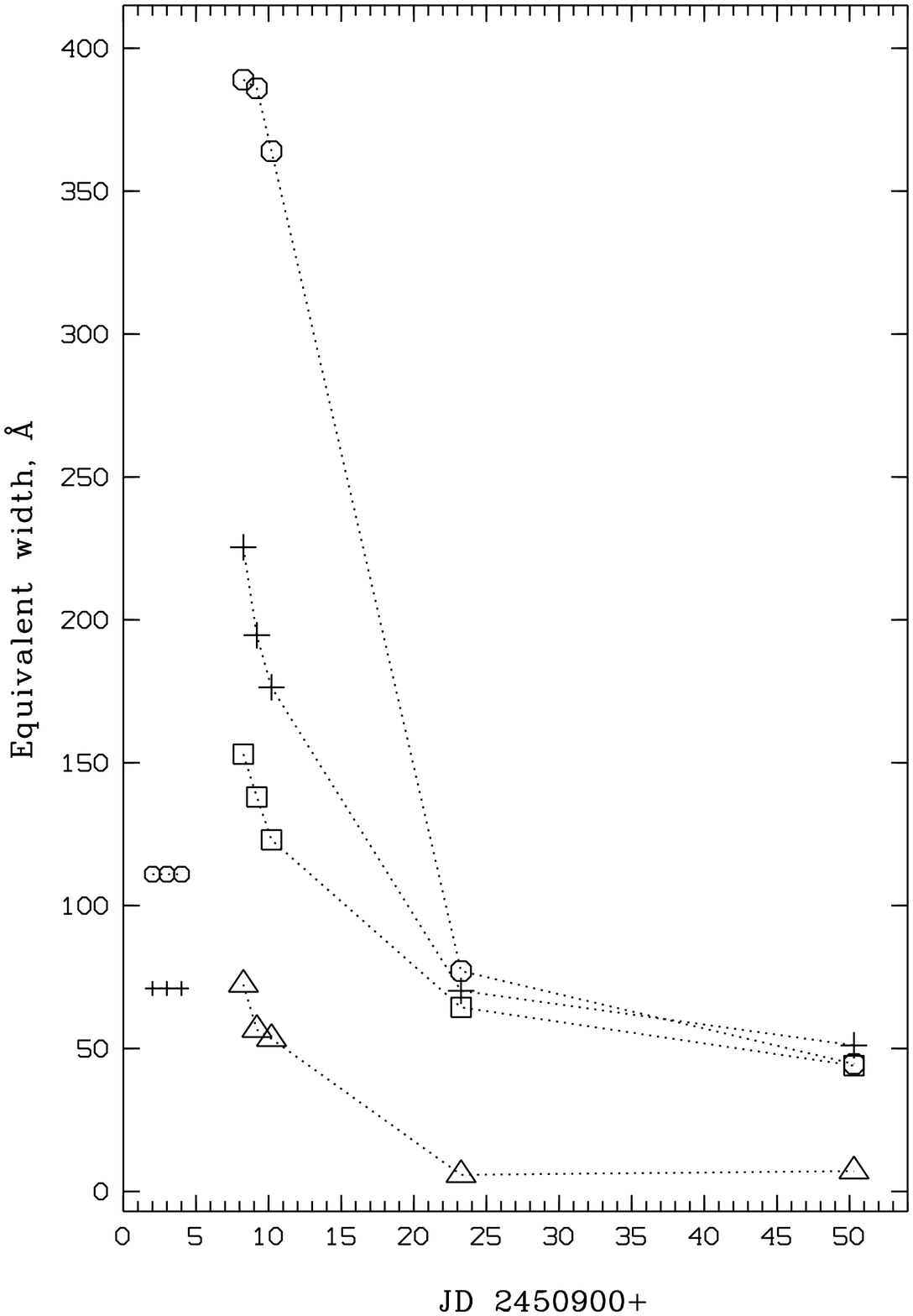,width=8cm}}
\caption{
Equivalent widths of the strongest emission lines.
HeI\,$\lambda$\,5876 (total profile)~-- circles,
H$\beta$\,(total)~-- crosses, H$\beta$\,(narrow)~-- squares,
H$\beta$\,(broad)~-- triangles. Small symbols show the equivalent widths
in  the quiescent state in January 1984 (Downes, 1984). The X--ray
maximum, as it follows from the RXTE observations, corresponds to  the
date JD\,2450904.5.}
\end{figure}
\begin{figure}
\centerline{\psfig{figure=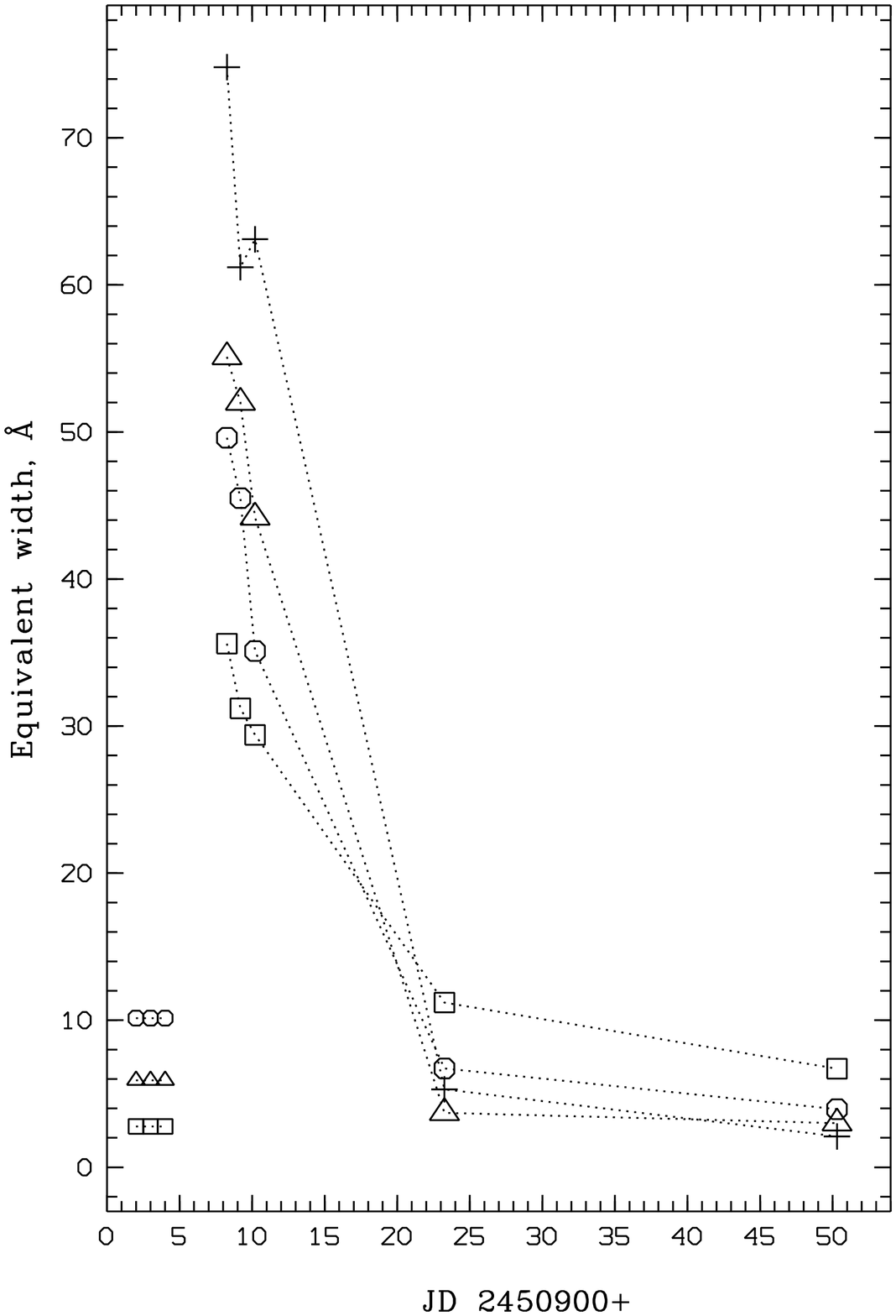,width=8cm}}
\caption{
 The same as Fig.~3 for emission lines: HeII~$\lambda\,4686$~-- crosses,
the blend SiII\,$\lambda 5041, 5056$, HeI\,$\lambda 5048$, CII\,$\lambda
 5045$ ~-- triangles, HeI~$\lambda\,4471$~--
circles, FeII~$\lambda5169$~-- squares.}
\end{figure}

Fig.\,2a,\,c demonstrate very bright and rich FeII spectrum.  The
intensities of FeII lines decrease with time, but the fading rate depends on
the line multiplet number and excitation conditions.  For example,
the FeII~$\lambda5169$ line, being  the strongest iron line just
after the burst, became weaker than the FeII~$\lambda 5317$ line  at the
end of the observing run.  Another example one can see in Fig.~2d,
where the intensity of the blend [NII]~$\lambda 5755$ + FeII~$\lambda 5748$
+ NII~$\lambda 5747$  even rises with time.

The line profile widths are observed to be different in  the lines
of different excitation.
Hydrogen and HeI lines have narrow and broad components. This is clearly
seen  on the strongest and unblended lines H$\beta$ and HeI~$\lambda
5876$ (Fig.~2). The
narrow component widths are about 400~km/s (corrected for the instrumental
line width)  at  the beginning of
observations,  and
they decrease to about 200 km/s  at the end of observations. The
broad components  decrease rapidly in intensity, but their
widths remain approximately the same, about 1200~km/s. FeII, TiII and
other lines all show only narrow components.

In our spectra we certainly see  some faint narrow absorption lines.
Nevertheless we do not confirm
the absorption lines of CaI, CoI, NiI, CrII and others reported by
Miroshnichenko (1995). It could be due
to real variability of the spectrum during the burst. Alternatively it
could be  due to the difference in the quality of the spectra which
we compare.
Our analysis of the absorption spectrum is still in progress.

\begin{figure}
\centerline{\psfig{figure=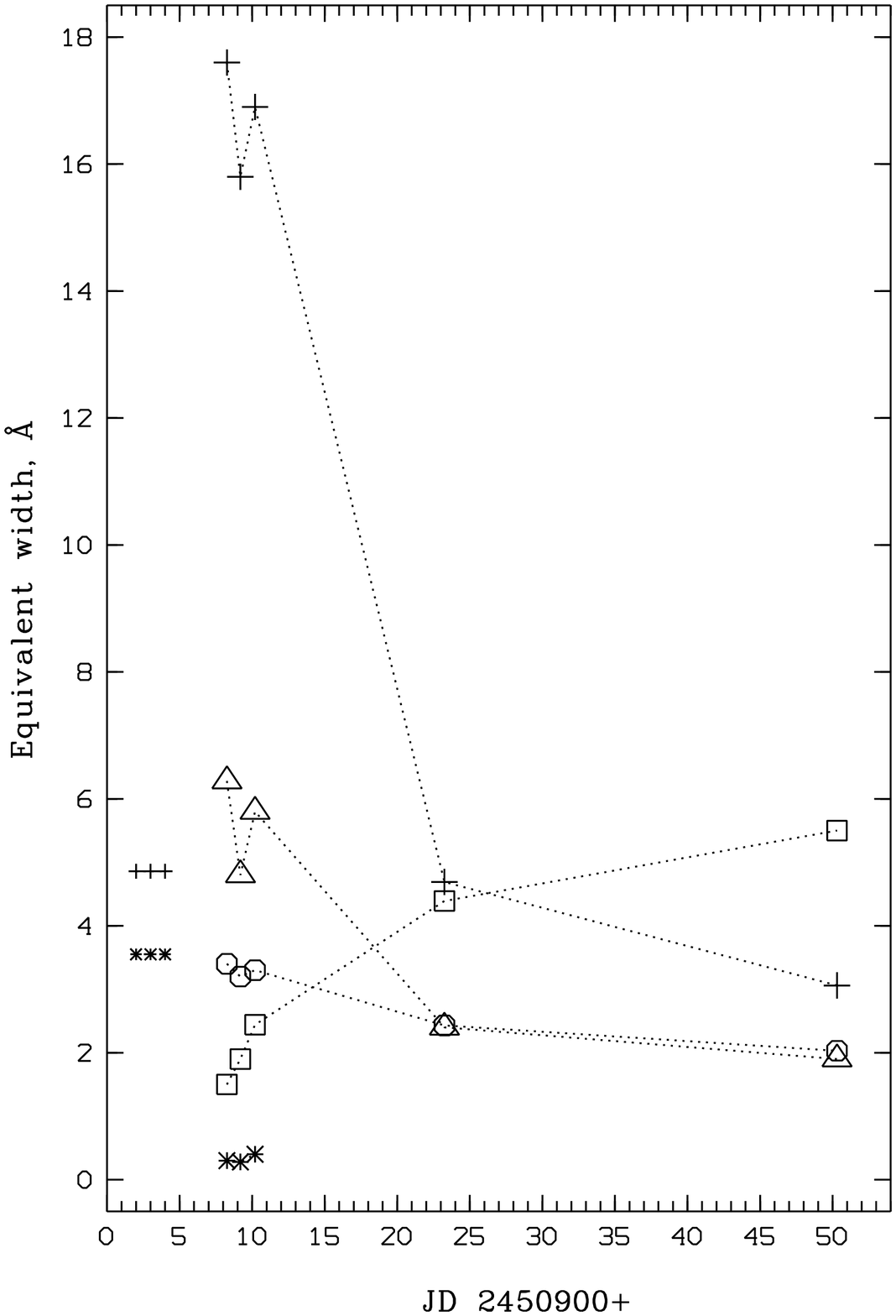,width=8cm}}
\caption{
  The same as Fig.\,3 for emission lines: the blend
ScII\,$\lambda 5658,\,5667,\,5669,\,5684$ + NII\,$\lambda 5676$ +
FeII\,$\lambda 5658$ -- crosses, MgI\,$\lambda\,5184$ -- triangles,
the blend CrII\,$\lambda$\,5502, 5503, 5509, 5511 +
FeII\,$\lambda\,5507$ -- circles, [NII]\,$\lambda\,5755$ -- squares,
weak blended lines FeII\,$\lambda 5748$ + NII\,$\lambda 5747$ -- asterisks.}
\end{figure}

The excitation level in the after-burst spectra  decreases with time.
In Fig.~3 we show the equivalent widths of total profiles of
H$\beta$ and HeI~$\lambda5876$ lines. In two weeks after the burst
 the H$\beta$ intensity became about the same as in  the quiescent
state of the spectrum observed by Downes (1984).  It is very interesting,
however, that the HeI intensity   dropped to a considerably
lower level than that observed in the quiescent state. This could be
a consequence of two different reasons:
\begin{enumerate}

\item The burst  broke down the line forming region (outer parts of
the accretion disk?).

\item The line forming region became more optically thick in these
particular lines because of the burst.
\end{enumerate}

 From  April 4 to 19 the strongest drops  were observed
in H$\beta$ broad  component (by a factor of 12.5) and in the SiII blend
(by a factor of 14.9, Fig.~4). Very strong fading  was
also observed  in the high excitation lines HeII (a factor of 14.1),
HeI~$\lambda\,4471$ (a factor of 7.4) and
HeI~$\lambda$\,5876 (a factor of 5.1). The H$\beta$ narrow  component (Fig.~3)
and FeII~$\lambda$\,5169 line (Fig.~4)  show a moderate drop (by factors of
2.4 and 3.0).

 We conclude that the X-ray burst resulted
both in creation of the broad line region with fast moving material
and in significant heating of the existing
narrow line region. The line broadening is probably  caused by Doppler
motion of emitting gas,  but not  by electron
scattering in the optically thick envelope. This is because of the fact that
on April 19 we observed both the broad  component fading and the drop
in helium lines (total profile) below
the standard quiescent level. For the same reason one can conclude that
the second scenario  of the two  mentioned above  is hardly probable.
So the outer parts of the accretion disk could be destroyed (or squeezed?)
because of the burst.

The strong drop in the intensity of SiII lines (Fig.~4) is not
easy to understand.
It could be connected with some fluorescent process.
We have mentioned above the remarkable behaviour of the [NII] emission line
whose intensity rises after the burst (Fig.~5). This indicates that
an extended envelope  has appeared. It is very important to search for
other forbidden lines in the CI\, Cam spectrum.
The [NII]~$\lambda5755$ line is blended in its blue wing with the lines
FeII~$\lambda 5748$ + NII~$\lambda 5747$. These two lines are very weak
in our spectra. Their intensity is 10 times less than that  obtained
from the spectrum of Downes (1984) and it slightly rises with time
after the burst.

All the data obtained from  the optical monitoring ---  emergence
of the high excitation spectrum, the broad line region,  and even
the  nebular forbidden line region are in good agreement with the
formation of relativistic jets
in the burst. But the main mechanism of the jet formation is still
unclear. It is important that the peak X-ray luminosity in 2--10 keV
region is $\sim 10^{37}$~erg/s at the adopted distance of 1~kpc.
It is well below the Eddington luminosity, which is necessary
for jet production in a supercritical accretion disk, similar to
that we directly observe in SS\,433
(Fabrika, 1997). An
alternative mechanism looks very promising --- the collimated
accretion--ejection process in  strong magnetic field of a neutron star.

The X-ray--radio--optical burst and the appearance of relativistic
S-shaped SS\,433-like jets make CI\,Cam to be a very  interesting star.
The emergence of high excitation spectrum in the burst implies
that this star is an unusual symbiotic-type X-ray binary, probably
like  GX\,1+4 (Chakrabarty and Roche, 1997).
It is very important to continue observations of CI\,Cam, which gives us
a chance to study the ``jet forming machine on holiday''.
\newpage
\begin{acknowledgements}
The work was supported by the RFBR grant N\,96-02-16396.
\end{acknowledgements}

\end{document}